\documentstyle[12pt]{article}
\normalsize

\def\gone{\gamma^{ (1)}}
\def\gtwo{\gamma^{ (2)}}
\def\gA{\gamma^{(A)}}
\def\gstarone{g^{* (1)}}
\def\gstartwo{g^{* (2)}}
\def\gstarA{g^{* (A)}}
\def\gB{\gamma^{(B)}}
\def\gstarB{g^{* (B)}}

\newcounter{sxn}
\def\sx#1{\addtocounter{sxn}{1}
\bigskip\medskip \goodbreak \noindent{\large\bf
\centerline{\thesxn.~~#1}} \nobreak \medskip}
\def\sxn#1{\sx{#1} }

\newcounter{axn}

\def\br{\begin{thebibliography}{abc}}
\def\er{\end{thebibliography}}

\date{}

\begin{document}
\bibliographystyle{unsrt}
\footskip 1.0cm
\thispagestyle{empty}

\setcounter{page}{0}
\begin{flushright}
UAHEP 9312\\
INFN-NA-IV-40/93 \\
 October 1993\\
\end{flushright}
\begin{center}{\LARGE POISSON LIE GROUP SYMMETRIES\\
         FOR THE ISOTROPIC ROTATOR\\}
\vspace*{10mm}
\vspace*{6mm}
{\large
   G. Marmo $^{1}$,
          A. Simoni $^{1}$,
          A. Stern $^{2}$ \\ }
\newcommand{\bc}{\begin{center}}
\newcommand{\ec}{\end{center}}
\vspace*{10mm}
 1){\it Dipartimento di Scienze Fisiche dell' Universit\`a di Napoli,\\
     Mostra d'Oltremare pad. 19, 80125 Napoli, Italy}.\\
 \vspace*{5mm}
 2){\it Department of Physics, University of Alabama, \\
 Tuscaloosa, Al 35487, USA.}\ec

\vspace*{5mm}

\normalsize
\centerline{\bf ABSTRACT}
We find a new Hamiltonian formulation of the classical
isotropic rotator where
left and right $SU(2)$ transformations are not canonical symmetries
but rather Poisson Lie group symmetries.
The system corresponds to the classical analog of a
quantum mechanical rotator which
possesses quantum group symmetries.
We also examine systems of two classical interacting rotators having
Poisson Lie group symmetries.
\newpage
\newcommand{\be}{\begin{equation}}
\newcommand{\ee}{\end{equation}}
\baselineskip=24pt
\newcommand{\ba}{\begin{eqnarray}}
\newcommand{\ea}{\end{eqnarray}}
\newcommand{\no}{\nonumber}

\sxn{Introduction}

Recently there has been interest in examining symplectic structures
which possess Poisson Lie group symmetries.
[1-7]
The interest is due in part to applications to classically
integrable systems and in part to the claim that
Poisson Lie group symmetries are the classical
analog of quantum group symmetries.\cite{avan,bab,TT}
To establish the connection
between Poisson Lie groups
and quantum groups
the quantization procedure known as
deformation quantization is utilized.\cite{bffls}
The study of classical systems with Poisson Lie group symmetries
may thus provide physical insight into the corresponding quantum
group invariant system.

Poisson Lie group transformations are implemented
on phase space via group multiplication, and
in general, they are not canonical transformations
as they need not preserve the
symplectic structure.  However, they are defined so that
invariance of the Poisson brackets follows once
the parameters of the group of
transformations are allowed to have certain nonzero Poisson brackets
with themselves.  Group multiplication is then said to correspond to
a Poisson map.

Symplectic structures
possessing Poisson Lie group symmetries have been constructed on
spaces known as $classical$ $doubles$.
[1-4,7]
Particle dynamics was not considered in these treatments,
nor were applications to physical particle systems.
In this article we write down particle Hamiltonians on the classical
double corresponding to $SL(2,C)$,
 and prove that the resulting dynamics is identical to
that of the isotropic rotator.
The Hamiltonians and the symplectic structures are parametrized by
a single parameter $\lambda$, and for a particular limiting
value of $\lambda$ one recovers the standard formulation.
Away from the limiting value, we cannot express the Hamiltonian
as the square of the angular momentum, and further
the angular momentum
does not satisfy the $SU(2)$ algebra.
The rotator does possess the usual
left and right $SU(2)$ symmetries.  However for $\lambda $ away from the
limiting value, we find that
these symmetries are not canonical.
Rather they are Poisson Lie group symmetries.
We thus claim that it is possible
to quantize the isotropic rotator
(using the method of deformation quantization)
so that the resulting system possesses
left and right quantum group symmetries.
We plan to study the quantization in a later paper.

Within the context of our example,
we verify a claim\cite{AT,raj} that the charges generating the
Poisson Lie group symmetries are group-valued;
For us they are valued in the $dual$ of $SU(2)$.
Because the charges are group-valued, some novel features arise
when we consider a system of two or more
identical isotropic rotators.  These features center around
the question of what is the analog of the
``total charge" for the system.  The answer cannot be
the sum of the individual charges
since the sum is not a group operation.
A more natural analog to the ``total charge" for
systems with Poisson Lie group symmetries
 is obtained by taking the group product, but this
definition is not unique
since the product does not in general commute.
Further, the transformations generated by such charges are not the same
as those generated by the total charge
in the corresponding standard canonical formalism.

The outline of this article is as follows:
In Sec. 2 we review the standard Hamiltonian formulation of
the classical rotator.  The alternative Hamiltonian formulations
on the classical
double $SL(2,C)$ are given in  Sec. 3.  The charges generating left
$SU(2)$ transformations are constructed in this section.
In Sec. 4 we use another coordinatization of $SL(2,C)$ to
construct the charges generating right
$SU(2)$ transformations.  Systems of two interacting
rotators are examined in Sec. 5.
Here we obtain four different interactions for the rotators,
each of which
posses Poisson Lie group symmetries, and each of which can be
thought of as deformations of the usual spin-spin interaction.

\newpage

\sxn{Standard Hamiltonian Formalism }

In the standard Hamiltonian formalism for
the classical rotator one has a set of
angular momenta $\ell_i$, $i=1,2,3$, which
satisfy the $SU(2)$ Poisson bracket algebra
\be
\{\ell_i,\ell_j\}=c^k_{ij}\ell_k \quad ,
\ee
$c^k_{ij}=\epsilon_{ijk}$ being the structure constants.
 To obtain the entire phase space
we must include the analog of position variables.
These variables indicate the orientation of the rotator.
We denote such variables by $g$ which here take
values in the group $G=SU(2)$.  The
phase space which results is known as
the cotangent bundle $T^*G$ of $G$.  In specifying the
Poisson brackets for $g$, one has
that the brackets of the components of $g$ (here represented by matrices)
with themselves are zero, while the Poisson brackets of
$\ell_i$ with $g$ are given by
\be
\{\ell_i,g\}=ie_i g \quad ,
\ee
$e_i$, $i=1,2,3$,
defining a basis for the Lie algebra ${\cal G}$ associated with $G$,
\be
[e_i,e_j]=ic^k_{ij}e_k \quad .
\ee
Eq. (2) was chosen so that $\ell_i$ generate left translations on $G$.
The latter are the canonical transformations corresponding to
spatial rotations.  Canonical transformations associated with
right translations on $G$ are generated by charges $t_i$ with
$t_i e_i=g^{-1} \ell_i e_i g$.

To determine dynamics for the system
we now specify the Hamiltonian.
The standard Hamiltonian for the isotropic rotator is
\be
H_0={1\over 2} \ell_i\ell_i\quad ,
\ee
where we have set the moment of inertia equal to one.  The resulting
system is rotationally invariant since $\{\ell_i,H_0\}=0$.
(It is also invariant under right $SU(2)$ translations
since $\{\ell_i,t_j\}=0$.)  Using eq. (4) the Hamilton equations
of motion for the system state are
\be
\dot \ell_i =0 \;    ,  \quad
\dot g g^\dagger =i \ell_i e_i  \quad .
\ee
Thus the angular momenta $\ell_i$
are constants of the motion, while $g$ undergoes a uniform precession.

\sxn{Alternative Hamiltonian Formalism }

We now give alternative Hamiltonian formulations of the isotropic
 rotator in which we modify i) the phase space,
ii) the nature of the symmetries and iii) the Hamiltonian.
Yet we preserve the dynamical system defined by the equations of motion
(5).
The alternative Hamiltonian formulations are all parametrized by
a single parameter $\lambda$, and for a particular limiting
value of $\lambda$ we recover the standard formulation described in
the previous section.

\medskip
i) We replace the phase space
$T^*G$ by a space known as the ${\it classical\;double}$ $D$
which we define below.
$D$ is a group which contains $G$
along with another subgroup
$G^*$,
which is the ${\it dual}$ of $G$.
$G^*$ has the same dimension of $G$.  Further,
let $e^i$, $i=1,2,3$, define a basis for the Lie algebra ${\cal G}^*$
associated with $G^*$, and $f_k^{ij}$ be the structure constants
for the algebra, ie.
\be
[e^i,e^j]=if_k^{ij}e^k \quad .
\ee
The Lie algebra ${\cal D}$ associated with $D$
is spanned by $e_i$ and $e^i$.
If for the Lie bracket between $e_j$ and $e^i$ one takes
\be
[e^i,e_j]=ic^i_{jk}e^k-if_j^{ik}e_k\quad ,
\ee
then there exists an invariant scalar product $<,>$
on ${\cal D}$ such that $e_i$ and $e^i$ are dual to each other, ie.
\be
<e_j,e^i>=<e^i,e_j>= \delta^i_j\;, \quad
<e_i,e_j>=<e^i,e^j>=0 \quad .
\ee
In this sense the group $G^*$ is ${\it dual}$ to $G$, and vice versa.
The algebra ${\cal D}$ is determined by the relations (3), (6) and (7).
The invariance of the scalar product $<,>$
is with respect to the adjoint action of $D$.  If $\alpha
\in {\cal D}$, then under the adjoint action an infinitesimal
change in $\alpha$ is given by
$\delta_\epsilon \alpha=[\epsilon,\alpha]$,
$\epsilon $ being an infinitesimal element of ${\cal D}$.
The invariance condition thus reads:
\be
<\delta_\epsilon \alpha ,\beta>+<\alpha,\delta_\epsilon \beta>=0\;,\quad
\alpha ,\beta \in {\cal D}\;.
\ee

For the case of interest where $G=SU(2)$,
the dual group $G^*$ is the group of $2 \times 2$ lower
triangular matrices with determinant equal to one.  It is
denoted by $SB(2,C)$.
The structure constants for $G^*$ can be chosen to be
$
f_k^{ij}=\epsilon_{ij\ell}\; \epsilon_{\ell k 3}\;.
$
In the defining representation for the algebra ${\cal G}$ and
${\cal G}^*$, the basis $e_i$ and $e^i$ satisfying
commutation relations (3), (6) and (7) can be
expressed in terms of Pauli matrices $\sigma_i$ according to:
\be
 e_i ={1\over 2} \sigma_i\;,\quad
 e^i ={1\over 2} (i\sigma_i +\epsilon_{ij3} \sigma_j)\;.
\ee
For the above representation
the scalar product $<,>$ defined in (8) corresponds to
$2$ times the imaginary part of the trace.
$e_i$ and $e^i$ together
span the $SL(2,C)$ Lie algebra.  Thus after exponentiation
we have that the classical double $D$ is $SL(2,C)$.

We shall coordinatize the phase space
$D=\{\gamma\}$ with variables in $G$ and in $G^*$.  Thus let
$g \in G$ and $g^* \in G^*$.  An element $\gamma$
in $D$ is then labeled by $(g^*,g)$ and can be
defined by using the Iwasawa decomposition
 $\gamma=g^* g$.  The coordinates $(g^*,g)$ of course
do not globally cover $D$ as, for instance, $(1,1)$ and $(-1,-1)$
are both mapped to the identity in $D$.  Nevertheless, they serve as a
useful parametrization of a finite region of $D$.

For the Poisson brackets of $g$ and $g^*$ we propose the following
quadratic\cite{SK} relations
\ba
\{g_1,g_2\}&=&[\;r^*_{12}\;,\;g_1 g_2\;] \;,
 \\
\{g_1^*,g_2^*\}&=&-[\;r_{12}\;,\;g_1^* g_2^*\;] \;,
\\
\{g_1^*,g_2\}&=&-g_1^*\; r_{12} \;g_2\;,
 \ea
where we use tensor product notation.
The indices $1$ and $2$ refer to two separate
vector spaces on which the matrices act.
$r_{12}$ and $r^*_{12}$
act nontrivially on both vector spaces $1$ and $2$,
while $g_1=g\otimes 1,\;g_2=1\otimes g,\;
g^*_1=g^*\otimes 1,\;g_2^*=1\otimes g^*.$
Antisymmetry for Poisson bracket relations (11) and (12) is insured
upon assuming the following conditions for matrices
$r_{12}$ and $r^*_{12}$:
\be
r_{12}^*=-r_{21}\quad {\rm and}\quad
r_{12}^* -r_{12}= {\rm adjoint}\;{\rm invariant} \;,
\ee
where adjoint invariant means:
$\gamma_1 \gamma_2 ({\rm adjoint}\;{\rm invariant})=
({\rm adjoint}\;{\rm invariant}) \gamma_1 \gamma_2\;,$
for any $\gamma \in D$.  Assuming these relations,
antisymmetry for the remaining Poisson bracket (13) implies that
\be
\{g_1,g_2^*\}=-g_2^*\; r_{12}^*\; g_1\;.
\ee
Jacobi identities involving $g$ and $g^*$
are satisfied provided the $r$ matrices fulfill
two quadratic equations,
\be
[r_{13},r_{12}]+[r_{23},r_{12}]+[r_{23},r_{13}] = 0 \;,
\ee
\be
[r_{23},r_{31}]+[r_{31},r_{12}]+[r_{12},r_{23}]  =
{\rm adjoint}\;{\rm invariant} \;.
\ee
(along with the relations obtained by interchanging the three vector
spaces.)
Eq. (16) is the classical Yang-Baxter equation, while (17) is a modified
classical Yang-Baxter equation.
We have used conditions (14) to derive (16) and (17).
Two solutions to (16) and (17) are:
\be
r_{12}=\lambda\; e^i \otimes e_i
\ee
or
\be
r_{12}=\mu \;e_i \otimes e^i \;,
\ee
where $\lambda$ and $\mu$ are constants.
{}From $r_{12}$ and eqs. (14) we can obtain $r^*_{12}$.

{}From (11)-(13) we can compute Poisson brackets for
elements $\gamma$ in $D$.  Using $\gamma=g^* g$, we find
\be
\{\gamma_1,\gamma_2\}=-\gamma_1\gamma_2\;r^*_{12}-r_{12}\;\gamma_1
\gamma_2\; \;.
\ee
This symplectic structure has been studied in \cite{AM}.  In \cite{AM}
it was shown that using solution (18) the corresponding symplectic
two form is a deformation of the symplectic two form for $T^*G$.
In what follows, we shall be interested in only solution (18)
for this reason.

It is easy to show that symplectic structure defined by
(11)-(13) is a deformation
of the Poisson brackets for a rigid body.
The deformation parameter is $\lambda$.
It is clear from (11) and (18) that the brackets of components of $g$
with themselves are zero in the limit $\lambda \rightarrow 0$.  To
recover (1) and (2) from (12) and (13) in this limit,
one must also introduce the parameter
$\lambda$ in $g^*=g^*(\lambda)$.  We define $g^*(\lambda)$ as follows:
\be
g^*(\lambda)=e^{i\lambda \ell_i e^i} \;.
\ee
Upon expanding $g^*(\lambda)$ in $\lambda$ and keeping the first order
terms in eq. (13), we obtain
\be
i \lambda e^i \otimes \{ \ell_i , g \} = - \lambda e^i \otimes e_i g \;,
\ee
which is equivalent to eq. (2).  By keeping second order
terms in eq. (12), we obtain
\be
-\lambda^2 e^i \otimes e^j \{\ell_i,\ell_j\} =
-i\lambda^2 \ell_k [\;e^i \otimes e_i\;,\; e^k \otimes 1 +1\otimes e^k\;]
\ee
$$
=-\lambda^2 \ell_k c^k_{ij} \; e^i \otimes e^j \;,
$$
which is equivalent to eq. (1).

\medskip
ii)  One feature of the symplectic structure defined by (11)-(13) is
the existence of Poisson Lie group transformations.
In general Poisson-Lie group transformations
are not canonical transformations as they need not preserve the
symplectic structure.  However,
the symplectic structure can be made to be invariant
under such transformations if we let the parameters
of the transformations have nonzero Poisson brackets
with themselves.

Among transformations of this type for our system are the
right transformations of $G$ on
$D=\{\gamma\}$,
\be
\gamma\rightarrow \gamma h \;,\; h \in G
\ee
and the left action of $G^*$ on $D=\{\gamma\}$,
\be
\gamma\rightarrow h^* \gamma  \;,\; h^* \in G^*\;.
\ee
In terms of the coordinates $(g^*,g)$ this implies
\be
g\rightarrow gh\;, \quad g^*\rightarrow g^* \;,
\ee
for the former and
\be
g\rightarrow g\;, \quad g^*\rightarrow h^* g^*  \;,
\ee
for the latter.  By themselves
transformations (26) and (27) do not preserve the Poisson brackets
(11)-(13).  But (11)-(13) can be made to be invariant
under (26) if we insist that $h$ has the following Poisson bracket
with itself
\be
\{h_1,h_2\}=[\;r^*_{12}\;,\;h_1 h_2\;] \;,
\ee
and zero Poisson brackets with $g$ and $g^*$.
Then $SU(2)$ right multiplication is a Poisson map and (26) corresponds
to a Poisson Lie group transformation.
For (27) to be a Poisson Lie group transformation, $h^*$
must have the following Poisson bracket with itself
\be
\{h_1^*,h_2^*\}=-[\;r_{12}\;,\;h_1^* h_2^*\;] \;,
\ee
and zero Poisson brackets with $g$ and $g^*$.
Since the right-hand-sides of (28) and (29) vanish
in the limit $\lambda \rightarrow 0 $,
the transformations (26) and (27) become canonical in the limit.

We note that Poisson brackets (11)-(13) are invariant under the
simultaneous action of both $G$ and $G^*$ via (26) and (27).
For this we assume that
\be
\{h_1^*,h_2\}=0\;.
\ee
By comparing with eq. (13) we
conclude that the algebra of the observables
$g$ and $g^*$ is different from the algebra of the symmetries
parametrized by $h$ and $h^*$.

If we consider the infinitesimal version of the right action of $G$,
then from eq. (28) we find that the Poisson bracket algebra of the
corresponding infinitesimal parameters is isomorphic to ${\cal G}^*$.
Similarly, from
 eq. (29) we find that the Poisson bracket algebra of the
infinitesimal parameters for left $G^*$ transformations
is isomorphic to ${\cal G}$.

Two additional
Poisson Lie group transformations exist for this system and
they too become canonical in the limit $\lambda \rightarrow 0 $.
They correspond to the left action of $G$ on
$D=\{\gamma\}$,
\be
\gamma\rightarrow f \gamma  \;,\; f \in G
\ee
and the right action of $G^*$ on $D$,
\be
\gamma\rightarrow \gamma f^*  \;,\; f^* \in G^*\;.
\ee
Because we decompose $\gamma$ with that an element $g^*$ of $G^*$
on the left and an element $g$ of $G$ on the right,
these transformations have a complicated
action on the coordinates $(g^*,g)$.
Below we shall examine only the former set of transformations, ie.
the left action of $G$ on
$D$, because as we shall show they are deformations of the
ordinary rotations of the rotator.

The infinitesimal version of (31) is given by variations $
\delta \gamma =i\epsilon^b e_b\; \gamma$.  It in turn
leads to the following variations $\delta$ on $g$ and $g^*$
\be
\delta g =i\epsilon^b {(ad \;g^*)_b}^a \;e_a g\;,\quad
\delta g^* =i\epsilon^b \biggl(e_b g^* -
{(ad \;g^*)_b}^a\;g^*  e_a \biggr)\;,
\ee
where $ ad \;g^*$ denotes an element of the
adjoint representation of $G^*$,
$$
{(ad \;g^*)_b}^a  =<g^* e^a {g^*}^{-1},e_b> \;,
$$
and $\epsilon^b$ are the infinitesimal parameters of the transformation.
Just as the infinitesimal parameters of the right action of $G$,
satisfied a Poisson bracket algebra
which was isomorphic to ${\cal G}^*$, the same must be true
of the infinitesimal parameters $\epsilon^a$ if
the symplectic structure defined in (11)-(13)
is to be invariant under (31).  More specifically,
\be
\{\epsilon^a,\epsilon^b\}=-\lambda f^{ab}_c\epsilon^c \;, \quad
\{\epsilon^a,g\}=\{\epsilon^a,g^*\}=0\;.
\ee
To show invariance
we note that the Leibniz rule does not apply for $\delta$
acting on a Poisson bracket.  For example, in computing
$\delta\{g_1,g_2\}=\{g_1+\delta g_1,g_2+\delta g_2\}-\{g_1,g_2\}$ we have
\be
\delta\{g_1,g_2\}=\{\delta g_1,g_2\}+\{g_1,\delta g_2\}
-\{\epsilon^b,\epsilon^d\} {(ad \;g^*)_b}^a {(ad \;g^*)_d}^c
 e_a \otimes e_c\; g_1g_2 \;.
\ee
The last term in (35) cannot be ignored since like the other terms
it is first order in $\epsilon^a$.  Using (34) we can then show that
the right-hand-side of (35) is equal to
$[\;r^*_{12}\;,\;\delta g_1 g_2 + g_1\delta g_2\;]$,
and hence (11) is invariant.

To show that transformations (33) are deformations of the ordinary
rotations of a rigid body
 we expand $g^*=g^*(\lambda)$
in powers of $\lambda$ around $\lambda=0$.
To lowest order, (33) reduce to
\be
\delta g \rightarrow i\epsilon^a e_a g
=\epsilon^a \{\ell_a, g\}\;, \quad
\delta \ell_a \rightarrow - c^c_{ab} \epsilon^b \ell_c
=\epsilon^b \{\ell_b, \ell_a\}\;,
\ee
where we used (1) and (2) to compute the Poisson brackets.
{}From eq. (34) the parameters $\epsilon^a$ have zero Poisson brackets
with everything when $\lambda\rightarrow 0$ and hence (33) are
canonical transformations in the limit.  Eqs. (36) show that the
limiting transformations are generated by the angular momenta $\ell_a$,
and therefore
correspond to rotations of the rigid body.

It has been noted\cite{AT,raj} that the charges which generate Lie
Poisson transformations take values in a group.
This will now be made evident for our example.
For arbitrary $\lambda$, we can express the
variations (33) according to
\ba
\delta g_1 &=& {i\over \lambda} \epsilon^a
<\{g_1,g_2^*\}{g_2^*}^{-1},1\otimes e_a>_2 \;,
\\
\delta g_1^*&=& {i\over \lambda} \epsilon^a
<\{g_1^*,g_2^*\}{g_2^*}^{-1},1\otimes e_a>_2\;,
\ea
where $<,>_2$ indicates that the scalar product is taken
on vector space $2$.  Eq's (37) and (38) reduce to (36) in the limit
$\lambda\rightarrow 0$.
{}From the chain rule it follows that any
function ${\cal F}={\cal F}( g^*, g)$
of the coordinates $ g^*$ and $ g$ undergoes the variation
\be
\delta {\cal F}_1 = {i\over \lambda} \epsilon^a
< \{ {\cal F}_1, g_2^*\}g_2^{*-1},1\otimes e_a>_2\;,
\ee
under the left action of $G$ on $D$,
where ${\cal F}_1={\cal F}\otimes 1$.
The variation of any function ${\cal F}$ on $D$ due to
the left action of $G$ can therefore be obtained by computing the
Poisson bracket of ${\cal F}$
with $g^*$.  In this sense $g^*$ can be
thought of as the charges generating the left action of $G$ on $D$.
These charges are valued in $ G^*$.  The charges generating
the right action of $G$ on $D$ also take values in $G^*$.
(We construct them in Sec. 4.)  Conversely, the charges associated with
the action of $G^*$ on $D$ take values in $G$.

\medskip
iii)  Just as the Hamiltonian $H_0$ on $T^*G$ [Cf. eq. (4)]
describing a rigid body
is invariant under both the left and right actions of $SU(2)$,
we can insist that the corresponding
``deformed" Hamiltonian $H(\lambda)$ on $D$ be
invariant under both the left and right actions of $SU(2)$.
Let us also insist that $H(\lambda=0)=H_0$.
One possible Hamiltonian consistent with these requirements is
\be
H(\lambda)={1\over {2\lambda^2}}( Tr \gamma \gamma^\dagger-2) =
  {1\over {2\lambda^2}}(Tr g^* {g^*}^\dagger-2) \;,
\ee
where we now take $ \gamma$ and $g^*$ to be matrices
in the defining representation of $SL(2,C)$ and $SB(2,C)$ respectively,
with the generators $e_i$ and $e^i$
given by (10).  For the Hermitean conjugates of the generators we have
\be
 {e_i}^\dagger =e_i\;,\quad
 {e^i}^\dagger ={1\over 2} (-i\sigma_i +\epsilon_{ij3} \sigma_j)\;.
\ee
 The 2 in parenthesis in (40) was subtracted so that (40) reduces
to the standard rigid body Hamiltonian (4) when $\lambda\rightarrow 0
$, ie. $H(\lambda=0)=H_0$.  Although $H(\lambda)$ is invariant
under either the left or right actions of $G$, it is not
invariant under either the left or right actions of the dual group
$G^*$.  Thus dynamics breaks the $SB(2,C)$ Lie Poisson
symmetry of the symplectic structure.

To obtain Hamilton's equations of motion for the system we need
the Poisson brackets of ${g^*}^\dagger$ with the dynamical variables
$g$ and $g^*$.
These brackets are not already determined by (11)-(13).  We can make
this more explicit by applying the two dimensional representation
for $SB(2,C)$,
parametrizing $g^*$ by a real and a complex number according to
$$ g^*=\pmatrix{y & \cr z & 1/y \cr}\;,\quad y\in {\bf R}\;,
\quad z\in {\bf C}\;. $$
Upon substituting this form
into (12) using (10) we can obtain the Poisson
bracket of for instance $y$ with $z$, $\{y,z\}= {{i\lambda}\over 2}yz$,
but the Poisson bracket of $z$ with its complex conjugate is
undetermined.  Therefore the Poisson brackets of the matrix components
${g^*}^\dagger$ with $g^*$ are not fully determined.

To specify
the Poisson brackets of ${g^*}^\dagger$ with the dynamical variables,
we note that the group property and the algebra are preserved under
$$g^*\rightarrow {{g^*}^\dagger}^{-1} \;,\qquad
e^a \rightarrow {e^a}^\dagger \;.  $$
We shall require that the Poisson brackets of $g$ and $g^*$
 are preserved under this mapping as well.
Applying the mapping to (12) and (13) we obtain
\ba
\{{{g_1^*}^\dagger}^{-1}\;,\;g^*_2\}&=&
-\lambda\;[\;{ e^i }^\dagger\otimes e_i\;,\;
{{g_1^*}^\dagger}^{-1} g_2^*\;]\;,
\\
\{{{g_1^*}^\dagger}^{-1}\;,\;g_2\}&=&
-\lambda \;{{g_1^*}^\dagger}^{-1}\;({ e^i}^\dagger \otimes e_i)\;g_2\;\;.
\ea
{}From these relations and
$\{ {{g^*}^\dagger}^{-1}{g^*}^\dagger\;,\;\cdot\;\}=0$ we easily obtain
that
\ba
\{{{g_1^*}^\dagger}\;,\;g^*_2\}&=&
\lambda\;\biggl(
\;{{g_1^*}^\dagger}( { e^i }^\dagger\otimes e_i)\; g_2^*\;-\;
g_2^*({ e^i }^\dagger\otimes e_i)\; {{g_1^*}^\dagger}\biggr)
\\
\{{{g_1^*}^\dagger}\;,\;g_2\}&=&
\lambda \;({ e^i}^\dagger \otimes e_i)\; {{g_1^*}^\dagger}g_2\;\;.
\ea

For the equation of motion for $g^*$ we must find that
it is a constant of the motion since
$g^*$ is the charge associated
with the $SU(2)$ left symmetry.  Hamilton's equation of motion gives
$$
\dot g^* =\{H(\lambda),g^*\}
= {1\over {2\lambda}}
\biggl(e_i g^*\;Tr[ g^* {g^*}^\dagger  ( { e^i}^\dagger- e^i)]
-g^* e_i \;Tr [ {g^*}^\dagger g^* ( { e^i}^\dagger- e^i)]\biggr)
$$
\be
=- {i\over {4\lambda}}
\biggl( \sigma_i g^*\;Tr[ g^* {g^*}^\dagger  \sigma_i]
-g^*  \sigma_i \;Tr [ {g^*}^\dagger g^*  \sigma_i]\biggr)\;\;,
\ee
since we may replace $ { e^i}^\dagger- e^i$ by
$-i \sigma_i$ in the defining representation of $SL(2,C)$.
But it is not hard to show that the right-hand-side of (46) is zero.
Since both $ g^* {g^*}^\dagger $ and $ {g^*}^\dagger g^*$ are Hermitean
we can express them as a linear combination of the Pauli matrices and
the unit matrix.
Upon substituting this form into (46) we get the desired result, ie.
\be
\dot g^* =0 \;.
\ee

For Hamilton's equation of motion for $g$ we get
\be
\dot g =\{H(\lambda),g\}
={1\over {2\lambda}} e_i g\;
Tr [ {g^*}^\dagger g^* ( { e^i}^\dagger- e^i)]
=-{i\over {4\lambda}} \sigma_i g\;Tr [ {g^*}^\dagger g^* \sigma_i]\;\;,
\ee
or
\be
\dot g g^\dagger=-{i\over{2\lambda}}\;(  {g^*}^\dagger g^*)_{tl}
\ee
where $A_{tl}$ denotes the traceless part of the matrix $A$, ie.
$A_{tl}= A -{ 1\over 2} Tr[ A]\; \times {\bf 1}_{2 \times 2}$.

The right-hand-side of (49) is a traceless Hermitean
matrix.  From (47) it is also a constant
matrix.  Hence $g$ undergoes a uniform precession, and {\it we obtain
the same dynamics as that of the isotropic rotator.}
This is despite the fact that the Hamiltonian and symplectic structure
of the deformed system defined in (11)-(13)
differ from that of the standard Hamiltonian
formulation of the isotropic rotator.

The result that the system described here is isotropic is at first
sight surprising because if we expand the Hamiltonian (40) to
second order in $\lambda$, we get an anisotropic looking term
\be
H(\lambda)
={1\over 2} \ell_i\ell_i\;\biggl( 1 + {{\lambda^2}\over{12}}(\ell_3)^2
\biggr)+O(\lambda^4)\;\;.
\ee
But  $\ell_i$ in $g^*=g^*  (\lambda)$ cannot be interpreted as
the angular momenta of the rotator for the deformed system.  Rather,
\be
L_i=-{1\over {2\lambda}} Tr \sigma_i {g^*}^\dagger g^*
\ee
plays that role since we can rewrite the equations
of motion (47) and (49) according to
\be
\dot L_i =0 \;    ,  \quad
\dot g g^\dagger =i L_i e_i  \quad ,
\ee
which is identical to (5).
Unlike in the standard formalism we cannot express the Hamiltonian
$H(\lambda)$ as the square of the angular momentum.
 This is seen by comparing (40) with $L_iL_i$.  Further,
 $L_i=L_i(\lambda)$ does not
satisfy an $SU(2)$ algebra, that is
\be
\{L_i(\lambda),L_j(\lambda)\}\ne c^k_{ij} L_k(\lambda)
\ee
 (except in the
limit $\lambda \rightarrow 0$, where $L_i$ coincides with $\ell_i$).
To prove this it is sufficient to
consider the case of small (but not zero) $\lambda$.
By expanding $ {g^*}^\dagger g^*$ to second order in $\lambda$, we
find the following expression for $L_i(\lambda)$
\be
L_i(\lambda)=\ell_i - {\lambda \over 2}
 f^{ij}_k \ell_j \ell_k + O(\lambda^2) \;\;
=\ell_i + {\lambda \over 2}(\ell^2 \delta_{i3}-\ell_3 \ell_i)
 + O(\lambda^2) \;\;.
\ee
Now use eq. (1) to prove the result (53).  (For this we can show
that there are no corrections of
order $\lambda$ to the $SU(2)$ algebra
(1) coming from the expansion of $g^*(\lambda)$ in (12).)

As in the standard formulation of the isotropic rotator
the equations of motion
 are invariant under $SU(2)\times SU(2)$ transformations.
The latter were canonical transformations in the standard formulation,
whereas they
 correspond to Poisson Lie group symmetries for the formulation
presented here.   To see that
the equations of motion (47) and (49) are invariant under
transformations (24) and (31),
let us rewrite them as equations
for the $SL(2,C)$ group element $\gamma=g^* g$.  We get:
\be
\dot \gamma \gamma^{-1}=
-{i\over{2\lambda}}\;(\gamma \gamma^\dagger)_{tl} \;.
\ee
Eq. (55) is unchanged under
$\gamma\rightarrow f \gamma h,\;\; h,f \in SU(2)$.

\sxn{Alternative Parametrization of Phase Space}

In the preceding
Hamiltonian formalisms, the isotropic rotator
is invariant under both left and right $SU(2)$ transformations.
The left $SU(2)$ Poisson Lie group
transformations are generated by $g^*$, the
general form for infinitesimal transformations given by (39).
What are the generators of the right $SU(2)$ Poisson Lie group
transformations?

To answer this question it is easiest to introduce a new
parametrization of the phase space $D$.  It corresponds to
decomposing any element
 $\gamma\in D$ with an element $\tilde g$ of $G$
on the left and an element $\tilde g^*$ of $G^*$ on the right.
Thus locally we have
\be
\gamma=g^* g=\tilde g\tilde g^*    \;.
\ee
The symplectic structure on $D$ given by (20)
is recovered if for the coordinates $\tilde g$ and $\tilde g^*$
we take the following Poisson brackets
\ba
\{\tilde g_1,\tilde g_2\}&=&-[\;r^*_{12}\;,\;\tilde g_1 \tilde g_2\;] \;,
 \\
\{\tilde g_1^*,\tilde g_2^*\}&=&[\;r_{12}\;,\;\tilde g_1^*
\tilde g_2^*\;] \;,
\\
\{\tilde g_1^*,\tilde g_2\}&=&-\tilde g_2\; r_{12} \;\tilde g_1^*\;.
 \ea

In terms of the coordinates $\tilde g$ and $\tilde g^*$,
the left action of $SU(2)$ on $D$ now has a simple form
\be
\tilde g\rightarrow f\tilde g\;, \quad\tilde g^*\rightarrow\tilde g^* \;,
\ee
the symplectic structure (57)-(59) being invariant provided
\be
\{f_1,f_2\}=-[\;r^*_{12}\;,\;f_1 f_2\;] \;.
\ee
The infinitesimal form of (61) agrees with (34) upon setting
$f={\bf 1}+i\epsilon^a e_a$.

On the other hand,
right $SU(2)$ transformations have a more complicated action
on $\tilde g$ and $\tilde g^*$.  Infinitesimal variations are of the form
\be
\delta \tilde g =i\eta^b (ad \;{\tilde g^{*-1})_b}^a \;\tilde g
e_a \;,\quad
\delta \tilde g^* =i\eta^b \biggl(\tilde g^* e_b -
(ad \;{\tilde g^{*-1})_b}^a\;  e_a \tilde g^* \biggr)\;,
\ee
where $\eta^b$ are the infinitesimal parameters and,
in order for the symplectic structure be invariant, they satisfy
\be
\{\eta^a,\eta^b\}=\lambda f^{ab}_c\eta^c \;, \quad
\{\eta^a,\tilde g\}=\{\eta^a,\tilde g^*\}=0\;.
\ee
Eq. (63) is the infinitesimal version of (28) with
$h={\bf 1}+i\eta^a e_a$.
These variations are generated by $\tilde g^*$, as
 any function ${\cal F}={\cal F}(\tilde g^*,\tilde g)$
of the coordinates $\tilde g^*$ and $\tilde g$ undergoes the variation
\be
\delta {\cal F}_1 = {i\over \lambda} \eta^a
<\tilde g_2^{*-1} \{ {\cal F}_1,\tilde g_2^*\},1\otimes e_a>_2\;,
\ee
under the right action of $SU(2)$ on $D$.
Thus just as with the left action of $SU(2)$, the charges
$\tilde g^*$ generating
the right action take values in the dual group $SB(2,C)$.

Finally, we note that the invariant Hamiltonain (40) for the isotropic
rotator can be written solely in terms of the generators $\tilde g^*$
of right $SU(2)$ transformations,
\be
H(\lambda)=
  {1\over {2\lambda^2}}(Tr \tilde g^* \tilde g^{*\dagger}-2) \;.
\ee

\sxn{System of Two Rotators}

In the previous sections we examined the charges $g^*$ and $\tilde g^*$
which generate the left and right
$SU(2)$ Poisson Lie group symmetries of the isotropic rotator.
They were both valued in the dual of $SU(2)$.  In general,
Poisson Lie group symmetries are generated by group-valued charges.
\cite{AT}
When we put together two identical systems possessing Poisson Lie group
symmetries, what are the associated charges for the combined system?

For the case of two identical systems possessing canonical
symmetries the associated charges for the combined system
are sums of the charges for
the individual systems.  But clearly we cannot apply the same
procedure to the case of systems possessing
Poisson Lie group symmetries since the sum is not a group operation.
A more natural analog to the ``total charge" for
systems with Poisson Lie group symmetries
 is obtained by taking the group product.  However, such a
definition is not unique
since the product does not in general commute.   Thus
if we put together two isotropic rotators then there are two distinct
analogs to the ``total charge" associated with say left $SU(2)$,
and they generate two distinct sets of Poisson Lie group transformations.
Further, both of these transformations have the novel feature
 that the individual rigid bodies
are rotated by different amounts, and the rotation of one of
them depends on the coordinates of the other.
Thus neither set of transformations corresponds
to the rotations which would be generated by the total angular momentum
of the corresponding standard canonical formalism (although they all
coincide in the limit $\lambda\rightarrow 0$).
We show these results in what follows.

Say that $\gone$ and $\gtwo$ denote $SL(2,C)$ elements
corresponding to two distinct rigid rotators $1$ and $2$, while
and $\gstarone$ and $\gstartwo$
are the generators of the left $SU(2)$ Poisson Lie
groups for the two respective systems.  Thus, $\gA$ and $\gstarA$
satisfy the Poisson brackets:
\ba
\{{\gA}_1 , {\gA}_2   \}&=&- {\gA}_1 {\gA}_2
\;r^*_{12}-r_{12}\;  {\gA}_1 {\gA}_2 \; \; \;,   \\
\{{\gstarA}_1,{\gstarA}_2\}&=& -
[\;r_{12}\;,\;{\gstarA}_1 {\gstarA}_2\;] \;\;\;,   \\
\{{\gA}_1 , {\gstarA}_2   \}&=&-
\;r^*_{12}\;  {\gA}_1 {\gstarA}_2 \; \; \;, \quad {\rm for}\;A=1,2\;,
\ea
and
\be
\{ {\gA}_1 , {\gB}_2 \}=
\{ {\gstarA}_1 , {\gstarB}_2 \}=
\{ {\gA}_1 , {\gstarB}_2 \}=   0 \;\;,\quad {\rm for}\;A\ne B\;,
\ee
Generalizing formula (39), the left $SU(2)$ Poisson Lie group generated
by $\gstarA$ induces the variation
\be
\delta^{(A)} {\cal F}_1 = {i\over \lambda} \epsilon^a
< \{ {\cal F}_1, {\gstarA}_2\}{\gstarA}_2^{-1},1\otimes e_a>_2\;.
\ee
on an arbitrary function ${\cal F}$ of $\gone$ and $\gtwo$.
Upon identifying ${\cal F}$ with the dynamical variables $\gA$,
we then have that
\be
\delta^{(A)} \gA = i \epsilon^a e_a \gA\;\;,
\quad {\rm for}\; A=1,2\;,
\ee
and
\be
\delta^{(A)} \gB = 0\;\;,
\quad {\rm for}\; A\ne B\;,
\ee

Now define the product charge
$g^{*(12)}= \gstarone \gstartwo$.  Upon applying formula (69)
with $(A)=(12)$, it induces the following variations of $\gA$:
\ba
\delta^{(12)} \gone &=& i \epsilon^a e_a \gone\;\;,     \\
\delta^{(12)} \gtwo &=& i \epsilon^a
{(ad \;\gstarone)_a}^b   \;e_b \gtwo\;\;,
\ea
Thus rigid rotators $1$ and $2$ are transformed differently under
the action of $g^{*(12)}$.
Furthermore, since $\gstarone$ corresponds
to coordinates for rotator $1$, we find that the transformation of
rotator $2$ depends on the coordinates of rotator $1$.

Similarly, let us define the product charge
$g^{*(21)}= \gstartwo \gstarone$.  Transformations of
rotator $1$ induced by $g^{*(21)}$
depend on the coordinates of rotator $2$.  These
transformations are given by (73) and (74) with the indices $1$ and
$2$ interchanged.

If the dynamics for the system
is such that we have two noninteracting
isotropic rotators, ie. the Hamiltonian for the combined system
is just the sum of two free Hamiltonians $H^{(1)}$ and $H^{(2)}$,
\be
H^{(A)}=
  {1\over {2\lambda^2}}(Tr \gstarA {\gstarA}^\dagger-2) \;,
\ee
then charges $g^{*(12)}$ and $g^{*(21)}$ generate two distinct
Poisson Lie group symmetries.  When $\lambda\rightarrow 0$, these charges
have an identical limit and it just corresponds to the total angular
momentum for the combined system.  Thus only when $\lambda\rightarrow 0$,
do $g^{*(12)}$ and $g^{*(21)}$ generate the usual rotation symmetry
for the combined system.

Interactions can be introduced which break
one of the above Poisson Lie group symmetries, leaving the other intact.
For example, the interaction Hamiltonian term
\be
H^{(12)}=
  {1\over {2\lambda^2}}(
Tr g^{*(12)}{g^{*(12)}}^\dagger-2) \;,
\ee
is not invariant under $SU(2)$ transformations generated by $g^{*(21)}$.
On the other hand, (76)
is invariant under $SU(2)$ transformations generated by
$g^{*(12)}$.  To see this we only need to use the result that the
group product preserves the Poisson brackets, ie. it is a Poisson map.
Similarly, the interaction term
\be
H^{(21)}=
  {1\over {2\lambda^2}}(
Tr g^{*(21)}{g^{*(21)}}^\dagger-2) \;,
\ee
is invariant under $SU(2)$ transformations generated by
$g^{*(21)}$, but it is not
invariant under $SU(2)$ transformations generated by
$g^{*(12)}$.  Of course, neither (76)
or (77) are invariant under independent
left $SU(2)$ transformations of $1$ and $2$
generated by $\gstarone$ and $\gstartwo$,
respectively.

In the limit $\lambda\rightarrow 0$, both $SU(2)$ transformations
get identified and they are generated by the total angular momentum.
Furthermore, the Hamiltonian interactions $H^{(12)}$ and $H^{(21)}$
approach the same limit when $\lambda\rightarrow 0$.
For this set $\gstarA =\gstarA(\lambda)=
e^{i\lambda \ell^{(A)}_i e^i} \;$ and expand (76) and (77) to lowest
order in $\lambda$ to obtain
\be
H^{(12)}\;,\;H^{(21)}\; \rightarrow \;
{1\over 2} (  \ell^{(1)}_i  +\ell^{(2)}_i)^2   \;\;.
\ee
Thus in the limit $\lambda\rightarrow 0$, we recover
the usual spin-spin interaction from both $H^{(12)}$ and $H^{(21)}$.

In the above discussion we did not consider the possibility
of right $SU(2)$ Poisson Lie
group symmetries.  Eq. (26) showed that the coordinates $g^*$ for
a single rotator are unchanged by right $SU(2)$ transformations.
Therefore all of the Hamiltonians considered so far, including
the interaction Hamiltonians (76) and (77),
which are functions solely of $\gstarA$ and ${\gstarA}^\dagger$,
are invariant under
right $SU(2)$ transformations.
In Sec. 4 we denoted the generators of the right
$SU(2)$ Poisson Lie group symmetry for
a single isotropic rotator by $\tilde g^*$.
Let $\tilde\gstarone$ and $\tilde\gstartwo$
be the generators of the right $SU(2)$ Poisson Lie
groups transformations
corresponding to two distinct rigid rotators $1$ and $2$.
Then we can define another set of product charges
$\tilde g^{*(12)}=\tilde \gstarone \tilde\gstartwo$ and
$\tilde g^{*(21)}=\tilde \gstartwo \tilde\gstarone$ associated
with right transformations.  As with the corresponding left
transformations, the
rigid rotators $1$ and $2$ are rotated differently under
the action of $\tilde g^{*(12)}$ and $\tilde g^{*(21)}$, and
the transformation of one of the
rotators depends on the coordinates of the other.
 $\tilde g^{*(12)}$ and $\tilde g^{*(21)}$ generate symmetries for all
the previously discussed systems.  As before,
we can introduce new interactions which break
one of the symmetries and leaves the other intact.
In analogy with (76) and (77), such interaction terms are
\be
\tilde H^{(12)}=    {1\over {2\lambda^2}}
( Tr \tilde g^{*(12)}{\tilde g^{*(12)\dagger}}-2) \;,
\quad {\rm and } \quad
\tilde H^{(21)}=
  {1\over {2\lambda^2}}(
Tr \tilde g^{*(21)}{\tilde g^{*(21)\dagger}}-2) \;.
\ee
Now these interactions are invariant under left Poisson Lie group
transformations generated by $\gstarone$ and $\gstartwo$.  Furthermore,
they approach the same limit as did
$H^{(12)}$ and $H^{(21)}$ when $\lambda \rightarrow 0$, given in (78).

We have therefore found four different deformations
of the usual spin-spin coupling.  They correspond to the
Hamiltonian terms:
$H^{(12)}$, $H^{(21)}$, $\tilde H^{(12)}$ and $\tilde H^{(21)}$.
All of these interactions possess Poisson Lie group symmetries.

{\bf Acknowledgements}

 A. Stern was supported in part by the Department of Energy, USA under
contract number DE-FG-05-84ER-40141.
A. Stern wishes to thank the group in Naples
for their hospitality while this work was in progress.
We are grateful for discussions with S. Rajeev.
Without his invaluable assistance this work would not have
been possible.
We are also grateful for discussions with P. Vitale and I. Yakushin.

\end{document}